\documentclass[twoside]{article}
\usepackage{qic,epsfig}
\usepackage{amsmath}
\usepackage{graphicx}
\usepackage{esint}
\usepackage{subfigure}
\usepackage{dcolumn}

\begin{document}
\setlength{\textheight}{8.0truein}    

\runninghead{Title  $\ldots$}
            {Author(s) $\ldots$}

\normalsize\textlineskip
\thispagestyle{empty}
\setcounter{page}{1}

\copyrightheading{0}{0}{2003}{000--000}

\vspace*{0.88truein}

\alphfootnote

\fpage{1}

\centerline{\bf
DYNAMICS OF COUPLED CAVITY ARRAYS}
\vspace*{0.035truein}

\centerline{\bf EMBEDDED IN A NON-MARKOVIAN BATH}

\vspace*{0.37truein}
\centerline{\footnotesize
XINYU ZHAO\footnote{Email:xzhao1@stevens.edu}}
\vspace*{0.015truein}
\centerline{\footnotesize\it Center for Controlled Quantum Systems and Department of Physics
and Engineering Physics,}
\baselineskip=10pt
\centerline{\footnotesize\it  Stevens Institute of Technology, Hoboken,
New Jersey 07030, USA}
\vspace*{10pt}
\centerline{\footnotesize 
JUN JING}
\vspace*{0.015truein}
\centerline{\footnotesize\it Center for Controlled Quantum Systems and Department of Physics
and Engineering Physics,}
\baselineskip=10pt
\centerline{\footnotesize\it  Stevens Institute of Technology, Hoboken,
New Jersey 07030, USA}
\vspace*{10pt}
\centerline{\footnotesize 
J. Q. YOU}
\vspace*{0.015truein}
\centerline{\footnotesize\it Department of Physics and State Key Laboratory of Surface
Physics,}
\baselineskip=10pt
\centerline{\footnotesize\it   Fudan University, Shanghai 200433, China}
\vspace*{10pt}
\centerline{\footnotesize 
TING YU\footnote{Email:Ting.Yu@stevens.edu}}
\vspace*{0.015truein}
\centerline{\footnotesize\it Center for Controlled Quantum Systems and Department of Physics
and Engineering Physics,}
\baselineskip=10pt
\centerline{\footnotesize\it  Stevens Institute of Technology, Hoboken,
New Jersey 07030, USA}
\vspace*{0.225truein}
\publisher{(received date)}{(revised date)}

\vspace*{0.21truein}

\abstracts{
In this paper, the non-Markovian quantum dynamics of a coupled $N$-cavity
model is studied based on the quantum state diffusion (QSD) approach.
The time-local Di\'{o}si-Gisin-Strunz equation and the corresponding
exact master equation are derived for the model consisting of a coupled
cavity array. The resulting QSD equation serves as a stochastic solution
to a genuine $N$-partite continuous-variable (CV) system. Several
non-Markovian effects are studied in two interesting examples -- two-cavity
and three-cavity, under different boundary conditions. We have shown
that the environment-memory can facilitate the cat-like state transfer
from one cavity to another in the case of a strongly non-Markovian
environment. 
}{}{}

\vspace*{10pt}

\vspace*{3pt}
\communicate{to be filled by the Editorial}

\vspace*{1pt}\textlineskip    
\section{Introduction}

The dynamics of quantum open systems is an important research topic
in quantum optics, quantum dissipative system and quantum information
\cite{Openbook1,Openbook2}. Open system dynamics is also essential
for a deeper understanding of decoherence and implementation of quantum
control \cite{Paz-Zurek,Paz2008,Paz2009,HPZ,ESD,Halliwell_Yu,Zhang,An-Feng-Zhang,Hu2012,Zhang2010,Xiong2010,2HOGoan,2HO,Bellomo}.
While the theory of quantum open systems in a Markov regime is well
understood using the standard Lindblad master equations or the corresponding
Markov quantum trajectories \cite{Carmicheal,Dalibardetal,Gisin-Percival,Knight},
the recent surge of interest in non-Markovian phenomena requires a
set of systematic theoretical tools for solving quantum dynamic problems
in strong non-Markovian regimes. Among many recent research efforts
in developing a theoretical approach to the non-Markovian systems,
the non-Markovian quantum state diffusion (QSD) equation initially
proposed by Strunz and co-workers \cite{Diosi1,QSD} provides a very
powerful tool in dealing with quantum open systems coupled to a bosonic
environment. Remarkably, the Di\'{o}si-Gisin-Strunz QSD equation was
derived directly from an underlying microscopic model irrespective
of environmental memory time and coupling strength. Several physically
interesting models have been solved using the non-Markovian QSD approach
ranging from one qubit \cite{QSD2}, two-qubit \cite{XYZ}, $N$-qubit
systems \cite{QSD2011}, to multi-level systems \cite{Jing Jun,Nlevel}.
Furthermore, as a computational tool in practical applications, the
QSD approach has also showed its potential value in many interesting
problems including a theory of precise measurement \cite{Chen}, quantum
control dynamics \cite{JingPQ}, and quantum biology \cite{Eisfeld2011}.

Up to now, most extant research on the non-Markovian QSD equations
has been focused on discrete-variable systems such as qubits, qutrits,
spins etc. However, the non-Markovian QSD approach for multiple continuous-variable
(CV) systems is equally important since many CV systems are known
to have important applications in quantum information processing (QIP)
such as quantum teleportation, quantum error correction, quantum imaging,
and quantum computing \cite{RMP2005}. Furthermore, the CV systems
are an integral part of quantum dissipative systems, non-equilibrium
statistical physics and theory of quantum decoherence \cite{Paz-Zurek}.

It should be noted that the systems composed of a single cavity or
two coupled cavities have been studied and the exact master equations
have been derived by using the path integral approach \cite{Zhang,Zhang2010,Xiong2010,2HO,An-Feng-Zhang}.
Recently, the N-harmonic oscillator case is also studied by using
the phase-space formalism \cite{NQBM,NQBM2,NQBM_Li}. The purpose
of this paper is to derive the non-Markovian QSD equation and the
corresponding master equation for a coupled $N$-cavity model. The
coupled $N$-cavity model is an important model in QIP due to its
relevance to the quantum communication network schemes \cite{RMP2005,Qnet}.
As an interesting application, we study several phenomena concerning
quantum coherence transfer (approximate) assisted by the environmental
memory. Specifically, in the case of two-cavity example, we show that
the memory effect of a highly non-Markovian environment can be used
to accomplish a transfer of the so-called Schr\"{o}dinger cat-like
state between the two cavities contained in the cavity chain. This
phenomenon exists only in the strongly non-Markovian environment.
Furthermore, in the case of the three coupled cavities model, we study
the cat-like state transfer and entanglement transfer for different
boundary conditions, and show that the dynamics of the system is strongly
dependent on the boundary conditions \cite{Wuetal}.

The paper is organized as follows: In Sec.~\ref{model}, we derive
the exact QSD equation and the corresponding master equation for the
coupled $N$-cavity model. Then, in Sec.~\ref{Ex_2cav}, we study
a special example, the coupled two-cavity case. We show that the cat-like
state can be transferred approximately from cavity 1 to cavity 2 without
any direct couplings. We show that the cat-like state transfer can
be accomplished only for a strongly non-Markovian environment. Sec.~\ref{Ex_3cav}
considers the coupled three-cavity example. We show that the quantum
dynamics (cat-like state or entanglement dynamics) will be significantly
modified by the boundary conditions. We conclude in Sec.~\ref{conclusion}.


\section{Coupled $N$-cavity in a finite temperature bath}

\label{model} 

We consider a coupled cavity array interacting with a common bosonic
bath at temperature $T$. The total Hamiltonian of the system plus
environment can be written as, 
\begin{equation}
H_{\mathrm{tot}}=H_{\mathrm{s}}+H_{\mathrm{b}}+H_{\mathrm{int}},\label{Htot}
\end{equation}
where the Hamiltonian of the system is described by 
\begin{equation}
H_{\mathrm{s}}=\sum\limits _{i=1}^{N}\omega_{i}a_{i}^{\dagger}a_{i}+\sum\limits _{i=1}^{N}\lambda_{i}(a_{i}^{\dagger}a_{i+1}+a_{i}a_{i+1}^{\dagger}),\label{H_sys}
\end{equation}
where $\omega_{i}$ $(i=1,2,...N)$ are the frequencies for $N$ cavities,
and $\lambda_{i}$ are the coupling constants between $ith$ cavity
and $(i+1)th$ cavity. Two types of boundary conditions for this model
will be considered. The first one is the periodical boundary condition
(PBC), which means that $N$ cavities form a closed loop (i.e., $a_{N+1}=a_{1}$).
The other one is the open boundary condition (OBC). Namely, the cavity
chain has two open ends. We assume that all the cavities are coupled
to a common bath, which is composed of a set of harmonic oscillators
$b_{j}$, $b_{j}^{\dagger}$. 
\begin{equation}
H_{\mathrm{b}}=\sum\limits _{j}\nu_{j}b_{j}^{\dagger}b_{j}.\label{H_bath}
\end{equation}
The interaction Hamiltonian between the system and the environment
is $H_{\mathrm{int}}=\sum\limits _{i,j}(k_{ij}b_{j}^{\dagger}a_{i}+k_{ij}^{\ast}b_{j}a_{i}^{\dagger})$.
For simplicity, we assume that each $k_{ij}=g_{j}l_{i}$ is the product
of two coupling constants $g_{j}$ and $l_{i}$, then 
\begin{equation}
H_{\mathrm{int}}=\sum\limits _{j}(g_{j}b_{j}^{\dagger}L+g_{j}^{\ast}b_{j}L^{\dagger}),
\end{equation}
where the Lindblad operator $L=\sum\nolimits _{i=1}^{N}l_{i}a_{i}$.

In the case of the finite-temperature bath, we can transform the finite
temperature case into an effective zero temperature model by introducing
a fictitious bath \cite{QSD,TY2004}. The non-Markovian QSD equation
for the finite-temperature case is given by 
\begin{align}
\frac{\partial}{\partial t}\psi_{t}=(-iH_{\mathrm{s}}+Lz_{t}^{\ast}+L^{\dagger}w_{t}^{\ast})\psi_{t}-L^{\dagger}\int\nolimits _{0}^{t}ds\alpha_{1}(t,s)\frac{\delta\psi_{t}}{\delta z_{s}^{\ast}}-L\int\nolimits _{0}^{t}ds\alpha_{2}(t,s)\frac{\delta\psi_{t}}{\delta w_{s}^{\ast}},\label{FTQSD1}
\end{align}
where $z_{t}^{\ast}=-i\sum_{i}f_{i}z_{i}^{\ast}e^{i\nu_{i}t}$, $w_{t}^{\ast}=-i\sum_{i}h_{i}^{\ast}w_{i}^{\ast}e^{-i\nu_{i}t}$
are two statistically independent Gaussian noises with $f_{i}=\sqrt{\bar{n}_{i}+1}g_{i}$
and $h_{i}=\sqrt{\bar{n}_{i}}g_{i}$ as the effective coupling constants.
Here $\bar{n}_{i}=\frac{1}{\exp(\hbar\nu_{i}/k_{B}T)-1}$. The two
Gaussian noises satisfy the following relations, 
\begin{align}
M[z_{t}] & =M[z_{t}z_{s}]=0,\quad M[z_{t}z_{s}^{*}]=\alpha_{1}(t,s),\\
M[w_{t}] & =M[w_{t}w_{s}]=0,\quad M[w_{t}w_{s}^{*}]=\alpha_{2}(t,s).
\end{align}
where $\alpha_{1}(t,s)=\sum_{i}|f_{i}|^{2}e^{-i\nu_{i}(t-s)}$ and
$\alpha_{2}(t,s)=\sum_{i}|h_{i}|^{2}e^{i\nu_{i}(t-s)}$ are correlation
functions for the two effective baths.

In order to deal with the functional derivatives in Eq. (\ref{FTQSD1}),
It can be shown that the following two operators can be used to replace
the functional derivatives, $O_{1}(t,s,z^{\ast},w^{\ast})\psi_{t}=\frac{\delta\psi_{t}(z^{\ast},w^{\ast})}{\delta z_{s}^{\ast}}$
and $O_{2}(t,s,z^{\ast},w^{\ast})\psi_{t}=\frac{\delta\psi_{t}(z^{\ast},w^{\ast})}{\delta w_{s}^{\ast}}$,
satisfying the following equations, 
\begin{align}
\frac{\partial}{\partial t}O_{1}=[-iH_{\mathrm{s}}+Lz_{t}^{\ast}+L^{\dagger}w_{t}^{\ast}-L^{\dagger}\bar{O}_{1}-L\bar{O}_{2},O_{1}]-L^{\dagger}\frac{\delta}{\delta z_{s}^{\ast}}\bar{O}_{1}-L\frac{\delta}{\delta z_{s}^{\ast}}\bar{O}_{2},\label{FT_Eq_O1}\\
\frac{\partial}{\partial t}O_{2}=[-iH_{\mathrm{s}}+Lz_{t}^{\ast}+L^{\dagger}w_{t}^{\ast}-L^{\dagger}\bar{O}_{1}-L\bar{O}_{2},O_{2}]-L^{\dagger}\frac{\delta}{\delta w_{s}^{\ast}}\bar{O}_{1}-L\frac{\delta}{\delta w_{s}^{\ast}}\bar{O}_{2},\label{FT_Eq_O2}
\end{align}
where $\bar{O}_{i}=\int_{0}^{t}\alpha_{i}(t,s)O_{i}(t,s,z^{\ast},w^{\ast})ds$
($i=1,2$). According to Eqs.~(\ref{FT_Eq_O1}-\ref{FT_Eq_O2}),
the exact $O$ operators for the $N$-cavity model can be determined
as 
\begin{align}
O_{1}(t,s,w^{\ast}) & =\sum\limits _{i=1}^{N}p_{i}(t,s)a_{i}+\int_{0}^{t}q(t,s,s^{\prime})w_{s^{\prime}}^{\ast}ds^{\prime},\label{FTO1}\\
O_{2}(t,s,z^{\ast}) & =\sum\limits _{i=1}^{N}x_{i}(t,s)a_{i}^{\dagger}+\int_{0}^{t}y(t,s,s^{\prime})z_{s^{\prime}}^{\ast}ds^{\prime},\label{FTO2}
\end{align}
while the coefficients satisfy the following equations 
\begin{align}
 \frac{\partial}{\partial t}p_{i}(t,s)
 & = i\omega_{i}p_{i}(t,s)+i[\lambda_{i}p_{i+1}(t,s)+\lambda_{i-1}p_{i-1}(t,s)]\nonumber \\
 & +\sum\limits _{j=1}^{N}l_{j}^{*}p_{j}(t,s)P_{i}(t)+\sum\limits _{j=1}^{N}l_{i}p_{j}(t,s)X_{j}(t)-l_{i}Y(t,s),\label{ptpi}
\end{align}
\begin{align}
 \frac{\partial}{\partial t}x_{i}(t,s) & =-i\omega_{i}x_{i}(t,s)-i[\lambda_{i}x_{i+1}(t,s)+\lambda_{i-1}x_{i-1}(t,s)]\nonumber \\
 & -\sum\limits _{j=1}^{N}l_{i}^{*}x_{j}(t,s)P_{j}(t)-\sum\limits _{j=1}^{N}l_{j}x_{j}(t,s)X_{i}(t)-l_{i}Q(t,s),
\end{align}
\begin{equation}
\frac{\partial}{\partial t}q(t,s,s^{\prime})=\sum\limits _{j=1}^{N}l_{j}^{*}p_{j}(t,s)Q(t,s^{\prime}),
\end{equation}
\begin{equation}
\frac{\partial}{\partial t}y(t,s,s^{\prime})=-\sum\limits _{j=1}^{N}l_{j}x_{j}(t,s)Y(t,s^{\prime}),\label{ptyi}
\end{equation}
where $P_{i}(t)=\int_{0}^{t}\alpha_{1}(t,s)p_{i}(t,s)ds,$ $Q(t,s^{\prime})=\int_{0}^{t}\alpha_{1}(t,s)q(t,s,s^{\prime})ds,$
$X_{i}(t)=\int_{0}^{t}\alpha_{2}(t,s)x_{i}(t,s)ds,$ and $Y(t,s^{\prime})=\int_{0}^{t}\alpha_{2}(t,s)y(t,s,s^{\prime})ds,$
with the initial conditions $p_{i}(t,t)=l_{i}$, $x_{i}(t,t)=l_{i}^{*}$,
$q(t,t,s^{\prime})=y(t,t,s^{\prime})=0,$ $q(t,s,t)=-{\textstyle \sum\nolimits _{j=1}^{N}}l_{j}p_{j}(t,s),$
and $y(t,s,t)=\sum\nolimits _{j=1}^{N}l_{j}x_{j}(t,s).$

The coupled cavity model leads to two noise-dependent $O$ operators,
hence the derivation of the exact master equation from the exact QSD
equation is highly non-trivial. By using the Heisenberg approach \cite{QBM}
we can show that the exact master equation takes the following form,
\begin{eqnarray}
\frac{\partial}{\partial t}\rho &= -i[H_{s},\rho]+\{\sum_{ijk}l_{i}l_{k}^{*}F_{i,j}(t)[a_{j}\rho,a_{k}^{\dagger}]+l_{i}l_{k}^{*}G_{i,j}(t)[\rho a_{j},a_{k}^{\dagger}]\nonumber \\
& +l_{i}^{*}l_{k}U_{i,j}(t)[a_{j}^{\dagger}\rho,a_{k}]+l_{i}^{*}l_{k}V_{i,j}(t)[\rho a_{j}^{\dagger},a_{k}]+\mathrm{h.c.}\}.\label{MEQFTfinal}
\end{eqnarray}
where
\begin{eqnarray}
F_{i,j}(t) & = & \int_{0}^{t}\alpha_{1}(t,s)f_{i,j}(t,s)ds\nonumber \\
G_{i,j}(t) & = & \int_{0}^{t}\alpha_{1}(t,s)g_{i,j}(t,s)ds\nonumber \\
U_{i,j}(t) & = & \int_{0}^{t}\alpha_{2}(t,s)u_{i,j}(t,s)ds\nonumber \\
V_{i,j}(t) & = & \int_{0}^{t}\alpha_{2}(t,s)v_{i,j}(t,s)ds\label{FTF}
\end{eqnarray}
and the details of the derivation can be found in Appendix A. In a
special case that the bath is at zero-temperature $T=0$, $\bar{n}_{i}=0,$
the QSD equation reduces to 
\begin{equation}
\frac{\partial}{\partial t}\psi_{t}=(-iH_{\mathrm{s}}+Lz_{t}^{\ast}-L^{\dagger}\bar{O}_{1})\psi_{t},\label{QSD_0T}
\end{equation}
Here the corresponding $O$ operator becomes noise-independent, 
\begin{equation}
O_{1}(t,s)=\sum\limits _{i=1}^{N}p_{i}(t,s)a_{i},
\end{equation}
and the equations for the coefficients simplify as 
\begin{equation}
\frac{\partial}{\partial t}p_{i}(t,s)=i\omega_{i}p_{i}(t,s)+i[\lambda_{i}p_{i+1}(t,s)+\lambda_{i-1}p_{i-1}(t,s)]+\sum\limits _{j=1}^{N}l_{j}^{*}p_{j}(t,s)P_{i}(t),\label{dp}
\end{equation}
since $X_{j}(t)$ and $Y(t,s)$ are just zero in the case $T=0$.
Finally, the corresponding master equation can be derived without
using Heisenberg approach as 
\begin{equation}
\frac{d}{dt}\rho=-i[H_{\mathrm{s}},\rho]+\{\sum\limits _{i,j=1}^{N}l_{i}P_{j}^{\ast}(t)[a_{i},\rho a_{j}^{\dagger}]+\mathrm{h.c.}\}.\label{MEQ_0T}
\end{equation}

The exact master equation derived for the $N$-cavity model is valid
for an arbitrary finite temperature bath including the zero temperature
case. Markov limit is obtained when the correction function is represented
by a $\delta$-function. In the Markov limit, the time-dependent coefficients
in the exact master equation become constants. The equation reduce
to the standard Lindblad master equation \cite{Openbook1}. This model
serves as a good example showing that the QSD approach is a very useful
tool in deriving the non-Markovian master equation for an $N$-body
CV system. In the following two sections, we will discuss several
interesting examples with $N=2$ and $N=3$.


\section{Two coupled cavities}

\label{Ex_2cav} 

In this and the next sections, we will show several interesting numerical
results based on the simple examples $N=2$ and $N=3$. According
to the general QSD equation for $N$-cavity model in Sec.~\ref{model},
the zero-temperature non-Markovian QSD equation for the coupled two-cavity
model is obtained from Eq.~(\ref{QSD_0T}) by setting $L=a_{1}+a_{2}$
and $\bar{O}_{1}(t)=P_{1}(t)a_{1}+P_{2}(t)a_{2}$. Consequently, the
corresponding master equation is described by Eq. (\ref{MEQ_0T})
with the coefficients given by Eq.~(\ref{dp}).

Note that our derivations of the QSD and master equations are independent
on the choice of the environmental correlation functions $\alpha(t,s)$.
For simplicity, all the numerical simulations using the QSD equation
are based on the Ornstein-Uhlenbeck noise defined as $\alpha(t,s)=\frac{\gamma}{2}e^{-\gamma\left\vert t-s\right\vert }$.
Clearly, the Ornstein-Uhlenbeck process recovers the standard Markov
limit when $\gamma\rightarrow\infty$, $\alpha(t,s)\rightarrow\delta(t,s)$.
One of advantage of using the Ornstein-Uhlenbeck noise is that it
allows us to observe the crossover properties of quantum dynamics
from non-Markovian to Markov regimes. A small $\gamma$ typically
represents a non-Markovian noise with the finite memory time. Actually,
the Ornstein-Uhlenbeck can be also simulated by ``pseudo-mode''. For
example, in our two-cavity model, if only the second cavity is coupled
to a Markovian bath and the first cavity is coupled to the bath indirectly
via cavity 2 (i.e., $l_{1}=0$, $l_{2}\neq0$), it is equivalent to
the first cavity directly coupled to a non-Markovian bath with the
Ornstein-Uhlenbeck correlation function. This has been proved in Ref.
\cite{pseudomode}.

\begin{figure} [htbp]
\centerline{\epsfig{file=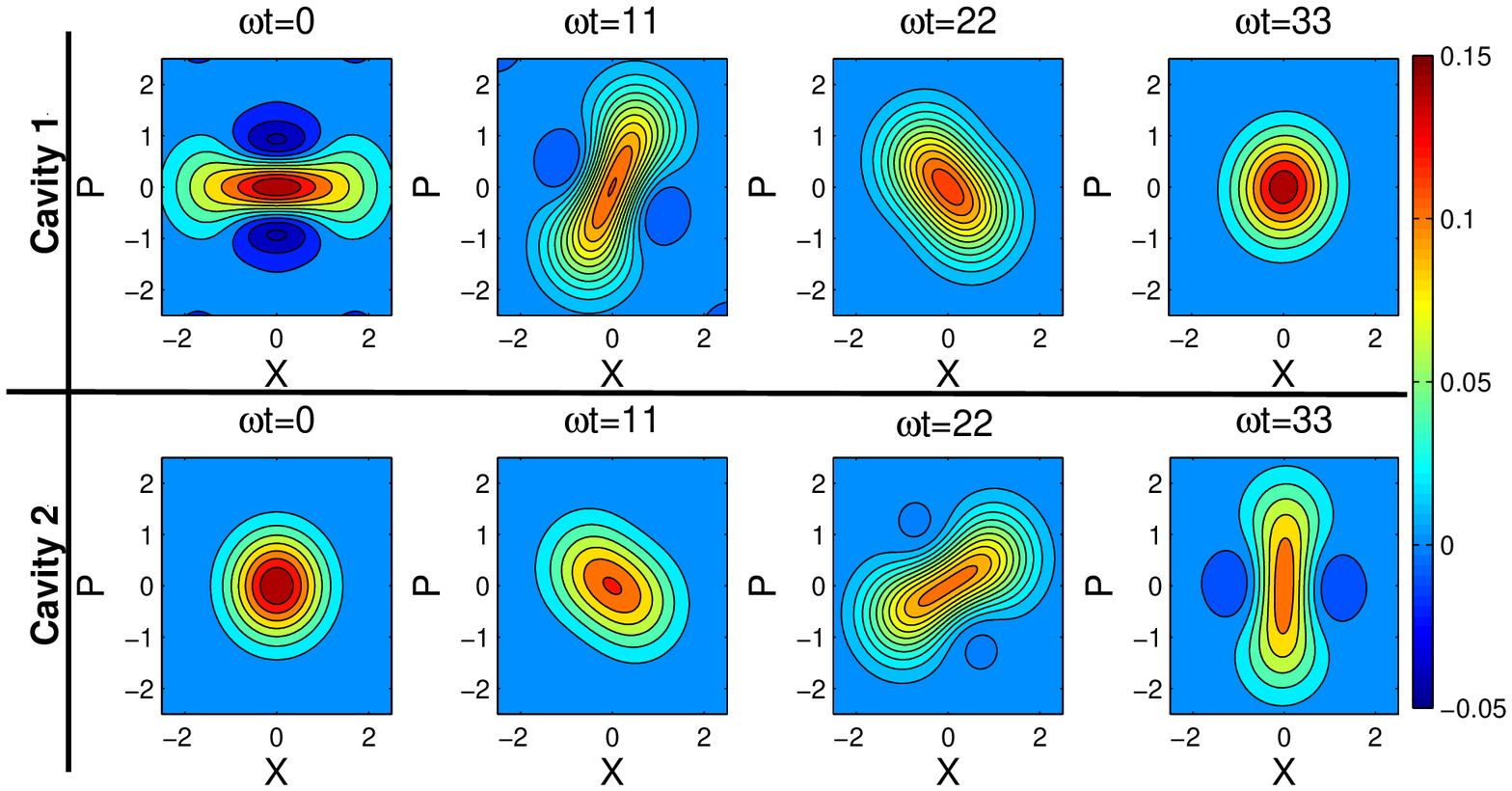, width=8.2cm}} 
\vspace*{13pt}
\fcaption{Wigner functions for the two cavities are plotted at
$\omega t=0$, $\omega t=11$, $\omega t=22$, and $\omega t=33$.
The initial state is chosen as the cat-state $1/\sqrt{Z}(\left\vert 1\right\rangle +\left\vert -1\right\rangle )\otimes\left\vert 0\right\rangle .$
The parameters are $\omega_{1}=\omega_{2}=\omega=1,$ $\lambda=0$,
$\gamma=0.1$.}
\label{cat_2cav}
\end{figure}

Obviously, if the two coupled cavities form a closed system, namely,
they are not coupled to an environment, one can see the state transfer
can be easily made through the mutual coupling between the two cavities.
In fact, if we choose the initial state of the first cavity as the
Schr\"{o}dinger's cat-like state $|\psi_{\mathrm{cat}}\rangle=1/\sqrt{Z}(|\alpha\rangle+|-\alpha\rangle)$
($Z$ is a normalization factor), while the second cavity is in the
vacuum state, i.e., $|\psi(0)\rangle=1/\sqrt{Z}(|\alpha\rangle+|-\alpha\rangle)_{1}\otimes|0\rangle_{2}$,
the cat-like state will be transferred to the second cavity from the
first cavity after time $t_{c}$: $|\psi(t=t_{c})\rangle=|0\rangle_{1}\otimes1/\sqrt{Z}(|\alpha\rangle+|-\alpha\rangle)_{2}$.
This result can be simply proved by solving the corresponding Fokker-Planck
equation \cite{ScullyQO}.

For the exact QSD equation or the master equation, we find that the
non-Markovian common bath can induce a mutual correlation between
two cavities indirectly, and make the cat-like state hop from one
cavity to another even the two cavities are not interacting to each
other directly. Notably, this phenomenon of the memory-assisted transfer
can be only observed in a strongly non-Markovian environment. It is
easy to show that the Markov environment will quickly suppress the
possibility of the state transfer due to the irreversible energy dissipation
to the environment. In Fig.~\ref{cat_2cav}, we plot the Wigner functions
for the two cavities at different time points. The regions with negative
values reflect the quantum interference due to the coherent superposition
of two wave packets. In a strongly non-Markovian case ($\gamma=0.1$),
the cat-like state can be transferred to the second cavity from the
first cavity without any direct coupling between two cavities (i.e.,
$\lambda=0$). As a comparison, it is noted that when the environment
is in a Markov regime ($\gamma$ is large), the high fidelity transfer
cannot happen (the Markov case is not plotted in the figure). Since
the dissipative process is dominant in the Markov limit, the decoherence
process suppresses the transfer process swiftly. Therefore, the cat-like
state may be destroyed by dissipation long before it can hop to the
second cavity. The numerical results show how a non-Markovian environment
affects the quantum state evolution.


\section{Three coupled cavities}

\label{Ex_3cav} 

A more complicated example is the case of three coupled cavities.
For $N=3$, the system Hamiltonian can be written as 
\begin{align}
H_{\mathrm{s}} & =\omega_{1}a_{1}^{\dagger}a_{1}+\omega_{2}a_{2}^{\dagger}a_{2}+\omega_{3}a_{3}^{\dagger}a_{3}+\lambda_{1}(a_{1}^{\dagger}a_{2}+a_{1}a_{2}^{\dagger})\nonumber \\
 & +\lambda_{2}(a_{2}^{\dagger}a_{3}+a_{2}a_{3}^{\dagger})+\lambda_{3}(a_{1}^{\dagger}a_{3}+a_{1}a_{3}^{\dagger}),
\end{align}
This model allows us to consider two types of boundary conditions.
In the case of the open boundary condition (OBC), we have $\lambda_{1}=\lambda_{2}=\lambda$,
$\lambda_{3}=0$. In contrast, in the case of the periodic boundary
condition (PBC), we have following constraints: $\lambda_{1}=\lambda_{2}=\lambda_{3}=\lambda$.
According to the QSD equation for the general $N$-cavity model discussed
in Sec.~\ref{model}, the exact QSD equation and the corresponding
master equation are given by Eq. (\ref{QSD_0T}) and Eq. (\ref{MEQ_0T}),
when $L=a_{1}+a_{2}+a_{3}$ and $\bar{O}_{1}=P_{1}(t)a_{1}+P_{2}(t)a_{2}+P_{3}(t)a_{3}$.
The coefficients can be determined by Eq. (\ref{dp}). The exact QSD
equation and master equation will serve as a powerful tool in our
study on the properties of the non-Markovian dynamics under two different
boundary conditions.

\begin{figure} [htbp]
\centerline{\epsfig{file=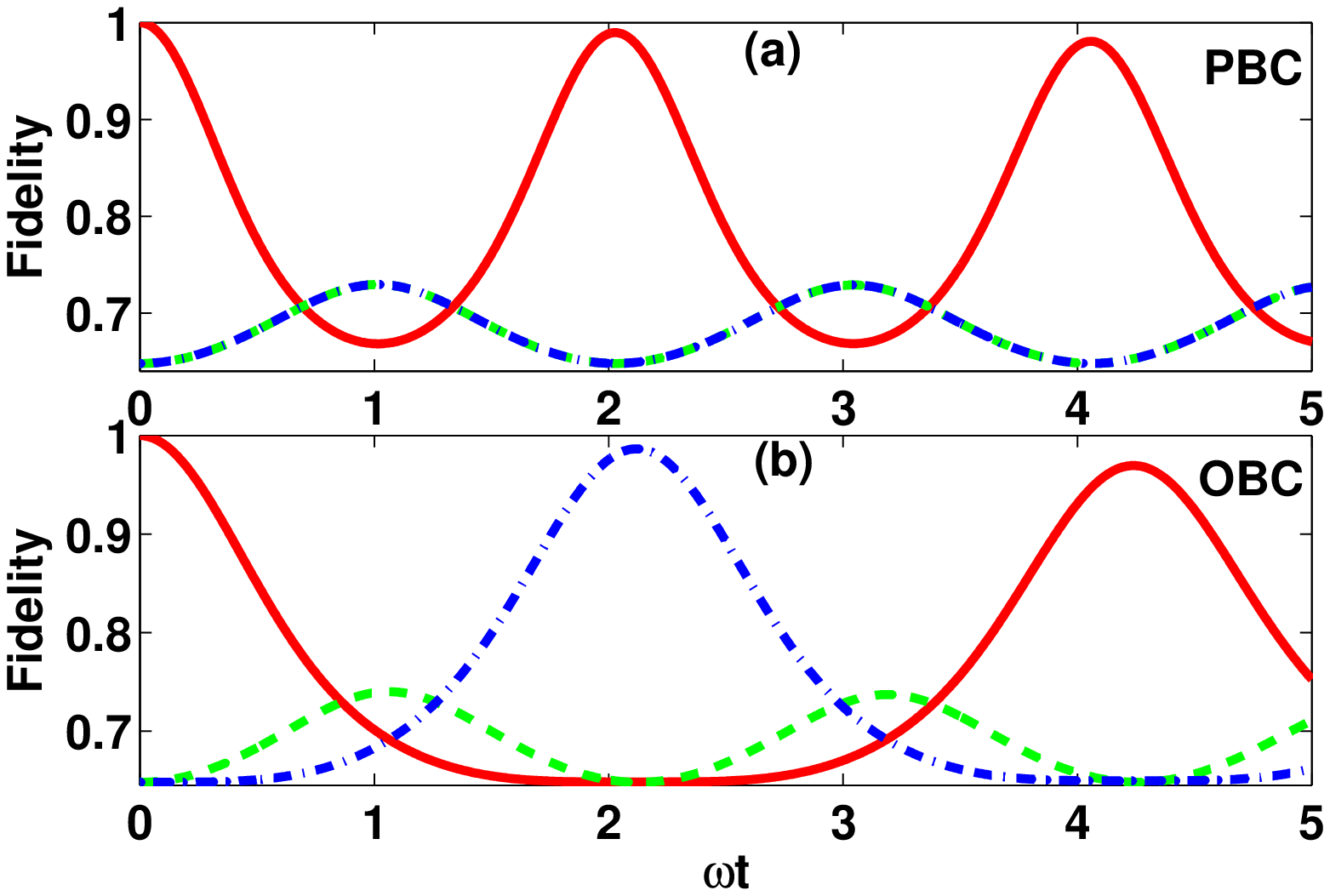, width=8.2cm}} 
\vspace*{13pt}
\fcaption{Time evolution of fidelities. (a) is for the case of
PBC, and (b) is for the case of OBC. The red (solid), green (dashed),
blue (dash-dotted) lines are the fidelities $\mathcal{F}^{(1)}$,
$\mathcal{F}^{(2)}$, and $\mathcal{F}^{(3)}$ for cavities 1, 2,
3, respectively. The parameters are $\omega_{1}=\omega_{2}=\omega_{3}=\omega=1$,
$\lambda=1$, $\gamma=0.2$.}
\label{fid_cat}
\end{figure}

\begin{figure} [htbp]
\centerline{\epsfig{file=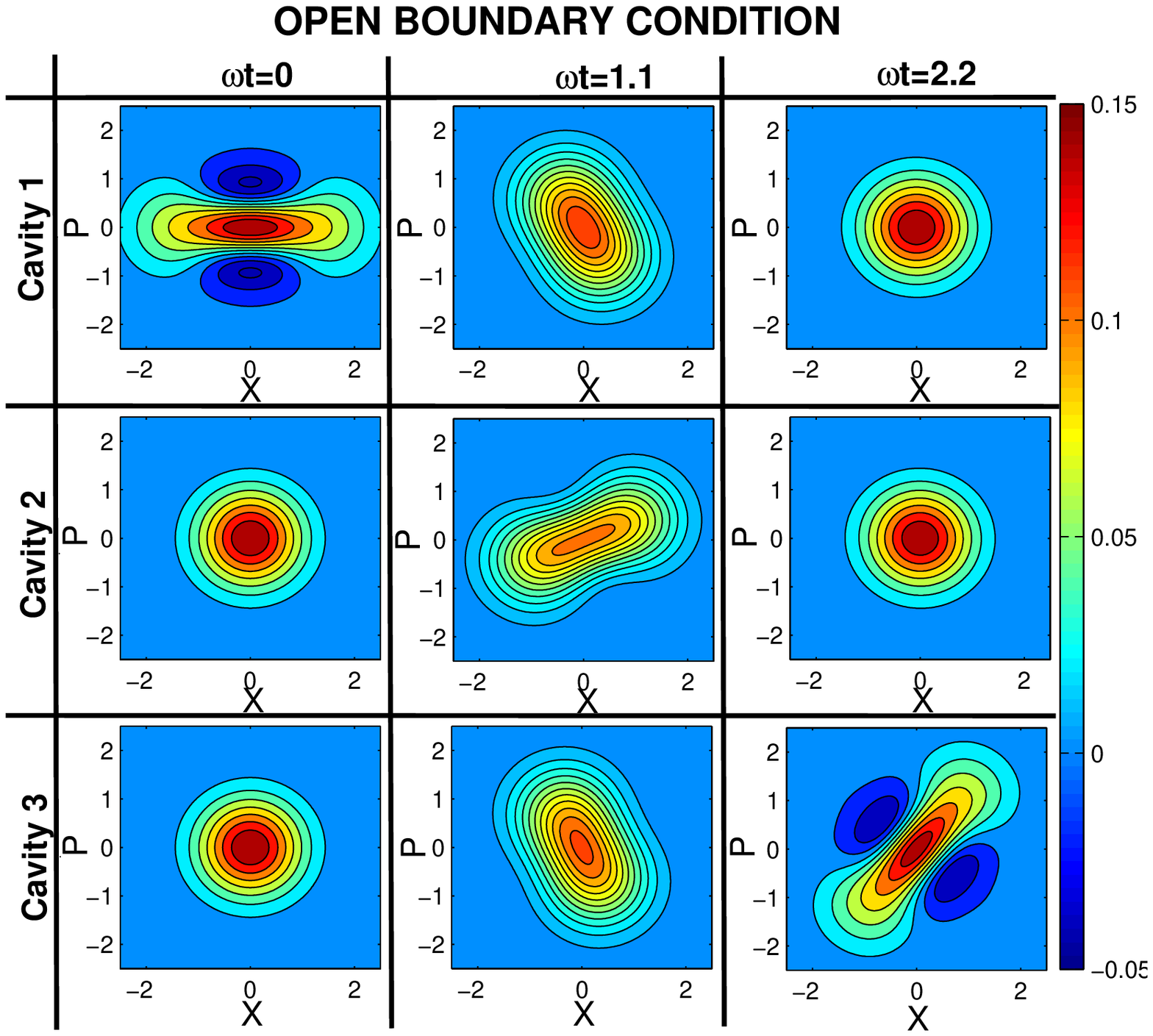, width=8.2cm}} 
\vspace*{13pt}
\fcaption{Wigner functions for three-cavity system with the OBC
at different time points. The initial state is chosen as $1/\sqrt{Z}(\left\vert 1\right\rangle +\left\vert -1\right\rangle )\otimes\left\vert 0\right\rangle \otimes\left\vert 0\right\rangle .$
The parameters are $\omega_{1}=\omega_{2}=\omega_{3}=\omega=1$, $\lambda_{1}=\lambda_{2}=1$,
$\lambda_{3}=0$, $\gamma=0.2.$}
\label{WFOBC}
\end{figure}

\begin{figure} [htbp]
\centerline{\epsfig{file=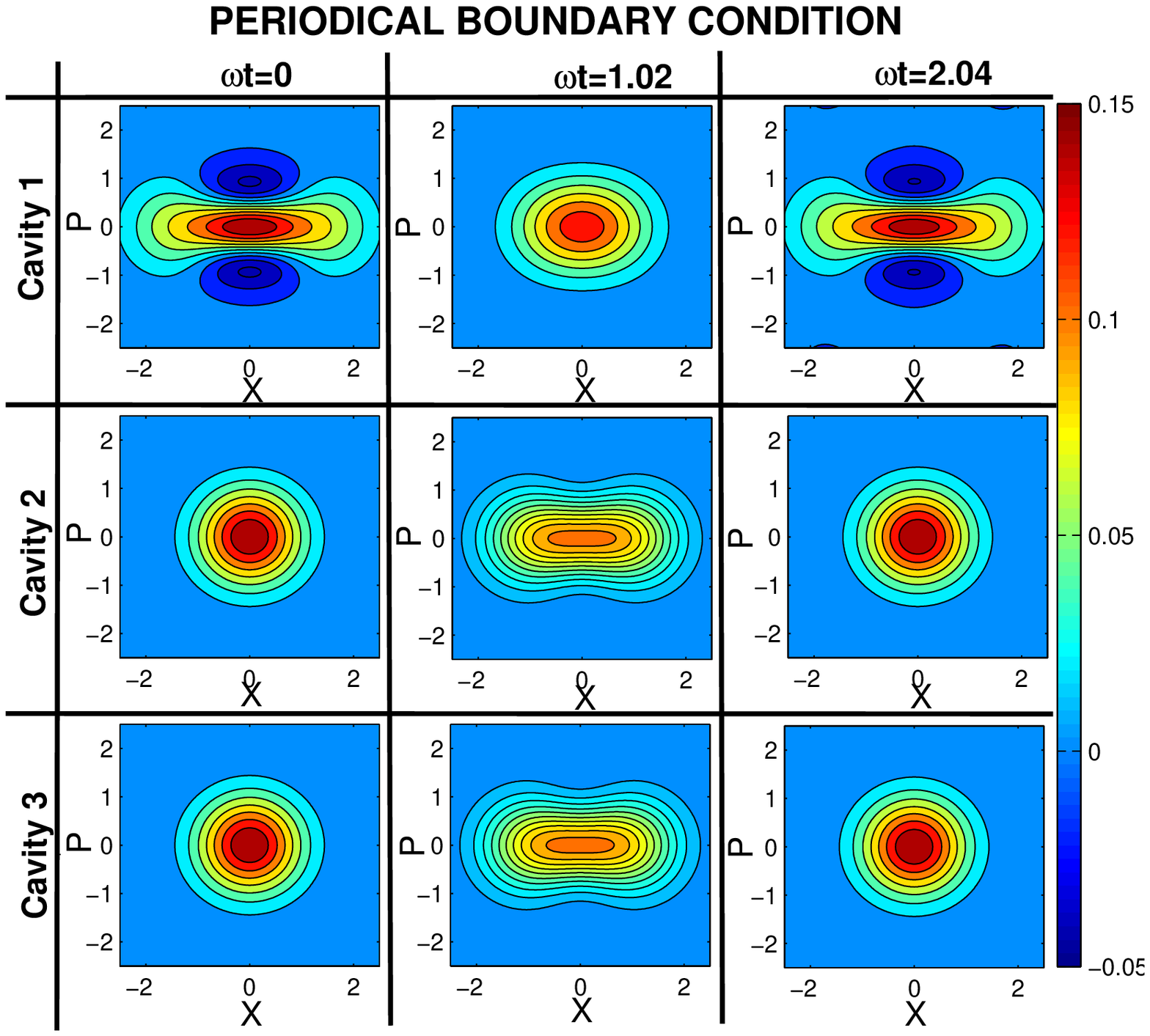, width=8.2cm}} 
\vspace*{13pt}
\fcaption{Wigner functions for three-cavity system with the PBC
at different time points. The initial state is chosen as $1/\sqrt{Z}(\left\vert 1\right\rangle +\left\vert -1\right\rangle )\otimes\left\vert 0\right\rangle \otimes\left\vert 0\right\rangle .$
The parameters are $\omega_{1}=\omega_{2}=\omega_{3}=\omega=1$, $\lambda_{1}=\lambda_{2}=\lambda_{3}=1$,
$\gamma=0.2.$}
\label{WFPBC} 
\end{figure}


\subsection{Cat-like state transfer under different boundary conditions}


We have already showed that the cat-like state $1/\sqrt{Z}(|\alpha\rangle+|-\alpha\rangle)$
can be transferred from one cavity to the other both analytically
and numerically in the case of $N=2$. However, the cat-like state
transfer can also be realized in the $N=3$ case. Moreover, the model
of three coupled cavities allows two different boundary conditions
that will profoundly affect the effect of state transfer. In Fig.~\ref{fid_cat},
fidelities are plotted for three cavities under two different boundary
conditions. The fidelities are calculated as $\mathcal{F}^{(i)}=\max_{\{\theta\}}[\langle\psi_{\mathrm{cat}}^{(R)}(\theta)|\rho^{(i)}(t)|\psi_{\mathrm{cat}}^{(R)}(\theta)\rangle]$,
where $|\psi_{\mathrm{cat}}^{(R)}(\theta)\rangle=1/\sqrt{Z}(|\alpha e^{-i\theta}\rangle+|-\alpha e^{-i\theta}\rangle)$
is the rotated cat-like state, $\rho^{(i)}(t)$ is the reduced density
matrix for the cavity $i$, and $\mathcal{F}^{(i)}$ is the fidelity
for the cavity $i$. Since the cat state may rotate in time evolution
no matter with or without an environment, so all the rotated cat-like
state $|\psi_{\mathrm{cat}}^{(R)}(\theta)\rangle$ for any given $\theta$
can be considered as a transferred cat-like state. Therefore, we calculate
the maximum fidelity between $\rho^{(i)}(t)$ and $|\psi_{cat}^{(R)}(\theta)\rangle$
for all possible $\theta$. From the curves of fidelities in Fig.~\ref{fid_cat},
it is shown that in the case of PBC, the fidelity will increase and
reach to 1 only in the first cavity. In the case of OBC, the fidelity
of the third cavity can be close to 1 at $\omega t\approx2.2$, which
means that the cat-like state is transferred into the third cavity.
It is worthwhile to point out that, in the case of OBC, the cat-like
state will only be transferred from cavity 1 to cavity 3. That is,
the state of the cavity 2 will never be a cat-like state. Intuitively,
in the case of OBC, two cavities at the end points of a chain are
symmetric while the middle cavity is in a unique position. In contrast,
for the case of PBC, the cat-like state never transfer into other
cavities since the other two cavities are in the symmetric positions.
The state has an equal probability to be one of the two cavities,
so none of two cavities cannot be in the cat state at the same time.
Therefore, one will expect that the cat state will experience a decay-revival
process in the initial cavity. To examine this more explicitly, in
Fig.~\ref{WFOBC} and Fig.~\ref{WFPBC}, we plot the Wigner functions
for each cavity in the three-cavity model with OBC and PBC. The time
evolution of the Wigner functions shows that the cat-like state can
transfer into cavity 3 at $\omega t=2.2$ in OBC case. However, in
the case of PBC, one only observes the decay and revival of the cat-like
state in the first cavity (initial cavity) at $\omega t=2.04$. Of
course, the cat state will disappear completely in a long-time limit
due to the dissipation effect.



\subsection{Entanglement transfer under different boundary conditions}

\begin{figure} [htbp]
\centerline{\epsfig{file=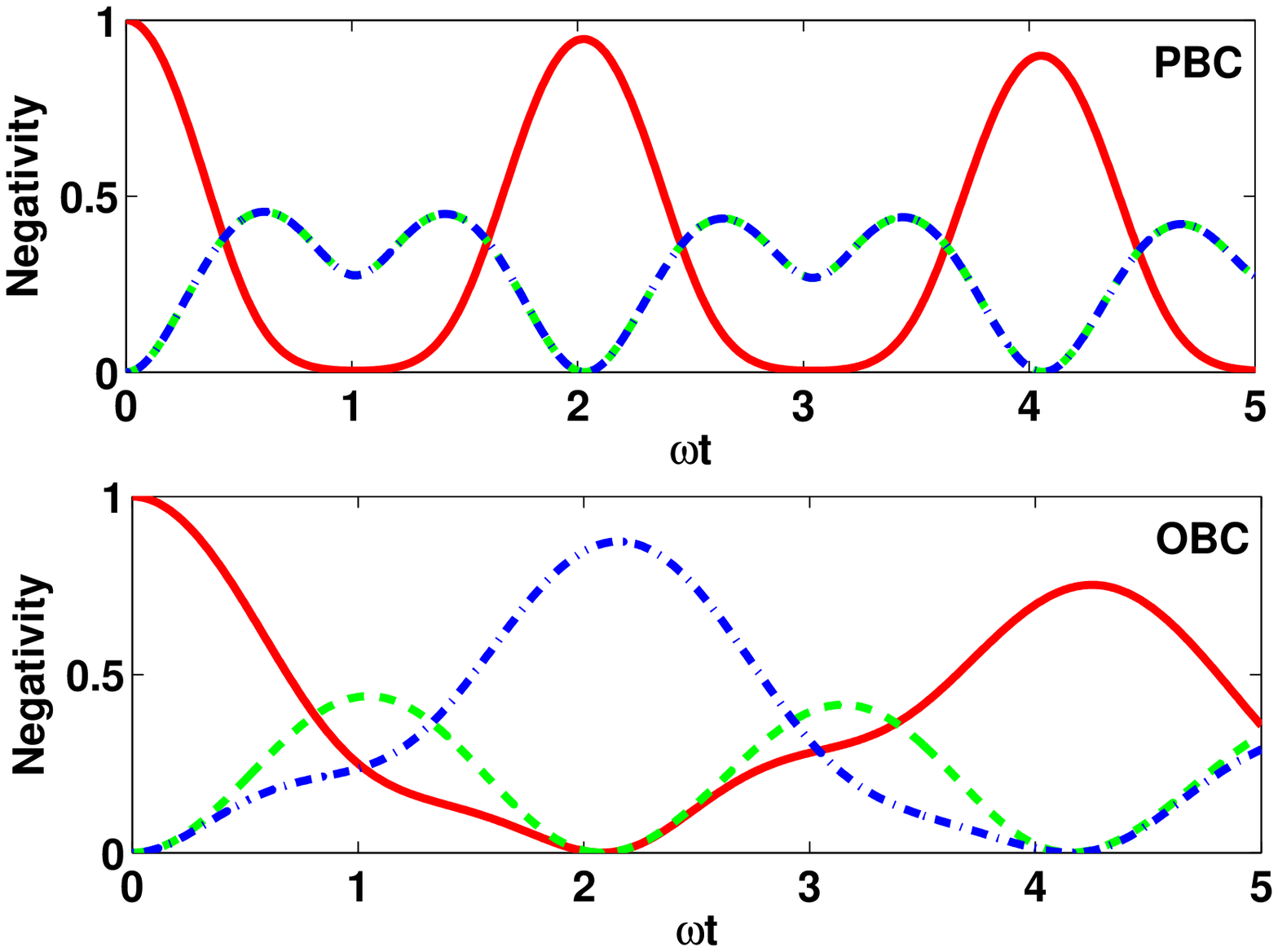, width=8.2cm}} 
\vspace*{13pt}
\fcaption{Time evolution of the negativity for all the three
modes. The red (solid), green (dashed) and blue (short-dashed) lines
are entanglement (negativity) between cavities 1-2, 1-3 and 2-3, respectively.
The initial state is chosen as $|\psi(0)\rangle=\frac{1}{\sqrt{2}}(|100\rangle+|010\rangle)$.
The parameters are $\omega_{1}=\omega_{2}=\omega_{3}=\omega=1,$ \ $\lambda=1,$
$\gamma=0.2.$}
\label{ent_trans} 
\end{figure}

\begin{figure} [htbp]
\centerline{\epsfig{file=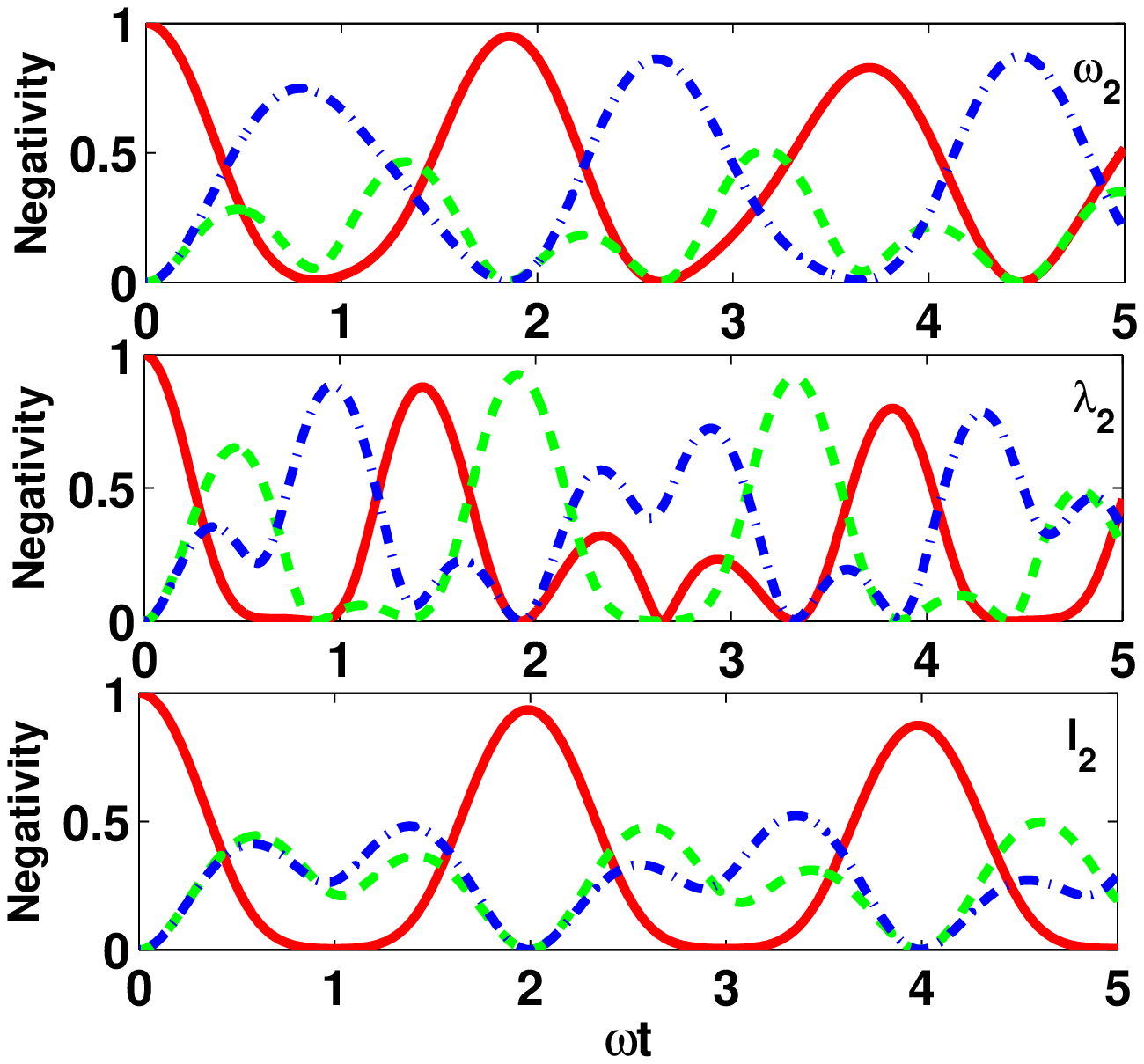, width=8.2cm}} 
\vspace*{13pt}
\fcaption{Symmetry breaking. The red (solid), green (dashed) and blue (short-dashed) lines
are entanglement (negativity) between cavities 1-2, 1-3 and 2-3, respectively. The parameters $\omega_{2}$, $\lambda_{2}$, and
$l_{2}$ are doubled in the sub-figures form top to bottom respectively.
The other parameters are the same as Fig. \ref{ent_trans} PBC case.}
\label{asymmetric}
\end{figure}

As another interesting phenomenon, let us consider the entanglement
transfer in the cavity array. We choose the initial state of the three-cavity
chain as $|\psi(0)\rangle=\frac{1}{\sqrt{2}}(|100\rangle+|010\rangle)$
(in Fock basis), where cavity 1 and 2 are entangled, but the combined
cavity 1 and 2 are separable from cavity 3. In the OBC case, the initial
entanglement between cavity 1 and 2 will transfer to the combined
system of cavity 2 and 3. Specifically, in Fig.~\ref{ent_trans}
(OBC case), the entanglement between cavity 2 and 3 will reach the
maximum value at $\omega t\approx2.2$. At the same time point, the
red line and green line in Fig.~\ref{ent_trans} show that cavity
1 become separable from both cavity 2 and cavity 3. As a comparison,
we see that, in the PBC case, the entanglement between cavity 2 and
3 never reach 1. However, PBC case is shown to be more helpful to
the entanglement preservation. For example, the loss of entanglement
degree in cavity 1 and 2 at $\omega t\approx4$ is less than that
in the case of OBC. In order to show the effect of asymmetric properties
on entanglement transfer, we break the symmetric by doubled the parameters
$\omega_{2}$, $\lambda_{2}$, and $l_{2}$. In Fig. \ref{asymmetric},
it is shown that the entanglement could also be transferred in PBC
case by breaking the symmetric properties of the parameters.

\section{Conclusion}

\label{conclusion}

In this paper, we have derived the time-local non-Markovian QSD for
the coupled $N$-cavity model embedded in a common bath without any
approximation. Using the non-Markovian quantum trajectory approach,
we have successfully derived the exact master equation for the $N$
coupled cavity model in both the zero and finite-temperature bath
cases. Based on these exact evolution equations, we studied several
interesting phenomena in the dynamic evolution of the two-cavity and
three-cavity cases. We studied the cat-like state transfer in the
coupled cavity array, and showed that a highly non-Markovian environment
can induce the cat-like state transfer (called the memory-assisted
cat state transfer). In the three-cavity example, we have studied
the effect of boundary conditions on the cat-like state transfer and
the entanglement transfer. These examples are used to show that the
non-Markovian QSD approach is a highly valuable tool in characterizing
the dynamics of a CV system. It is also important to note that the
stochastic approach used in this paper can be directly extended to
a generic multi-partite CV system with an arbitrary number of cavities.

\nonumsection{Acknowledgements}
\noindent
We acknowledge grant support from the DOD/AF/AFOSR No. FA9550-12-1-0001, The NBRPC No. 2009CB929300,
the NSFC Nos. 91121015 and the MOE No. B06011. J. Jing also acknowledges
grant support from NSFC No. 11175110, ``Chen Guang'' project by
Shanghai Municipal Education Commission and Shanghai Education Development
(No. CG0938) and Shanghai Leading Academic Discipline Project (No.
S30105)

\nonumsection{References}

\appendix

\nonumsection{Deriving the non-Markovian master equation in the finite temperature
case}
\noindent
\label{appendix A}
In this section, we will provide details of deriving the finite temperature
master equations from the exact QSD equations. Following the Heisenberg
approach \cite{QBM,TY2004}, one can deal with the derivation of the
exact master equation in the general case.

In the finite temperature case \cite{QSD,TY2004}, the total Hamiltonian
in the interaction picture is given by 
\begin{align}
H_{\mathrm{tot}}^{(I)}(t) & =H_{\mathrm{s}}+\sum_{i}(f_{i}^{\ast}e^{-i\nu_{i}t}L^{\dagger}d_{i}+f_{i}e^{i\nu_{i}t}Ld_{i}^{\dagger})+\sum_{i}h_{i}^{\ast}e^{-i\nu_{i}t}L^{\dagger}e_{i}^{\dagger}+h_{i}e^{i\nu_{i}t}Le_{i}),\label{FTHtot}
\end{align}
Using the Heisenberg evolution equation, we can derive the dynamic
equation for the following operators as 
\begin{align}
d_{i}(t) & =d_{i}-if_{i}\int_{0}^{t}L(s)e^{i\nu_{i}s}ds,\label{int_di}\\
e_{i}(t) & =d_{i}-ih_{i}^{\ast}\int_{0}^{t}L^{\dagger}(s)e^{-i\nu_{i}s}ds,\\
d_{i}^{\dagger}(s) & =d_{i}^{\dagger}(t)-if_{i}^{\ast}\int_{s}^{t}L^{\dagger}(s^{^{\prime}})e^{-i\nu_{i}s^{\prime}}ds^{\prime},\\
e_{i}^{\dagger}(s) & =e_{i}^{\dagger}(t)-ih_{i}\int_{s}^{t}L(s^{\prime})e^{i\nu_{i}s^{\prime}}ds^{\prime},\label{int_eid}
\end{align}

For a single trajectory, the time evolution of the state of the system
is governed by the stochastic propagator of the system $G(t,z^{\ast},w^{\ast})=\langle z|\langle w|U_{t}|0\rangle$
as $|\psi\left(t,z^{\ast},w^{\ast}\right)\rangle=G(t,z^{\ast},w^{\ast})|\psi_{\mathrm{s}}\left(0\right)\rangle$.
Applying Schr\"{o}dinger equation to the whole system and noticing
Eq.~(\ref{int_di}-\ref{int_eid}), we will find $G(t,z^{\ast},w^{\ast})$
satisfies an equation similar to the QSD\ equation (\ref{FTQSD1}),
\begin{align}
 \frac{\partial}{\partial t}G(t,z^{\ast},w^{\ast})=&-iH_{s}G+Lz_{t}^{\ast}G-L^{\dagger}\int_{0}^{t}ds\alpha_{1}(t,s)\langle z|\langle w|U_{t}L(s)|0\rangle\nonumber \\
 & +L^{\dagger}w_{t}^{\ast}G-L\int_{0}^{t}ds\alpha_{2}(t,s)\langle z|\langle w|U_{t}L^{\dagger}(s)|0\rangle.
\end{align}
From this dynamic equation, we can also define the $O$ operators
from the stochastic propagator 
\begin{equation}
\langle z|\langle w|U_{t}L(s)|0\rangle=O_{1}(t,s,z^{\ast},w^{\ast})G(t,z^{\ast},w^{\ast}),\label{def_O1}
\end{equation}
\begin{equation}
\langle z|\langle w|U_{t}L^{\dagger}(s)|0\rangle=O_{2}(t,s,z^{\ast},w^{\ast})G(t,z^{\ast},w^{\ast}).\label{def_O2}
\end{equation}
In the most general asymmetric case ($\omega_{i}$, $\lambda_{i}$,
$l_{i}$ are all different), we need to introduce many basis operators
$O_{1,i}$ and $O_{2,i}$ to expand $O$ operators as
\begin{equation}
\langle z|\langle w|U_{t}a_{i}(s)|0\rangle=O_{1,i}(t,s,z^{\ast},w^{\ast})G(t,z^{\ast},w^{\ast}),\label{def_O1i}
\end{equation}
\begin{equation}
\langle z|\langle w|U_{t}a_{i}^{\dagger}(s)|0\rangle=O_{2,i}(t,s,z^{\ast},w^{\ast})G(t,z^{\ast},w^{\ast}),\label{def_O2i}
\end{equation}
It is obviously that $O_{1}=\sum_{i}l_{i}O_{1,i}$, $O_{2}=\sum_{i}l_{i}^{*}O_{2,i}$.
Then, instead of deriving the non-linear differential equations of
$O$ operator with respect to $t$ such as Eq.~(\ref{ptpi}-\ref{ptyi}),
we will derive another set of differential equations with respect
to the initial value $s$, and it will be showed later that these
equations are linear. Taking the derivative with respect to $s$ of
Eq.~(\ref{def_O1i}), and noticing Eq.~(\ref{int_di}-\ref{int_eid}),
one gets,
\begin{eqnarray}
\frac{\partial}{\partial s}O_{1,i} & = & -i\omega_{i}O_{1,i}-i\lambda_{i}O_{1,i+1}-i\lambda_{i-1}O_{1,i-1}\nonumber \\
 &  & +w_{s}^{*}l_{i}^{*}-\sum_{j}l_{i}^{*}l_{j}\int_{0}^{s}ds^{\prime}\alpha_{1}(s,s^{\prime})O_{1,j}(t,s^{\prime},z^{\ast},w^{\ast})\nonumber \\
 &  & -\sum_{j}l_{i}^{*}l_{j}\int_{s}^{t}ds^{\prime}\alpha_{2}(s^{\prime},s)O_{1,j}(t,s^{\prime},z^{\ast},w^{\ast}).\label{psO1}
\end{eqnarray}
The same kind of equation for $O_{2,i}$ can be derived in a similar
way. Thus, we have obtained the differential equations for $O_{1,i}$
and $O_{2,i}$ with respect to $s$.

Recall the master equation derived in Ref.~\cite{TY2004}, 
\begin{align}
\frac{\partial}{\partial t}\rho=-i[H_{s},\rho]+[L,M\{P_{t}\bar{O}_{1}^{\dagger}\}]-[L^{\dagger},M\{\bar{O}_{1}P_{t}\}]+[L^{\dagger},M\{P_{t}\bar{O}_{2}^{\dagger}\}]-[L,M\{\bar{O}_{2}P_{t}\}].\label{MEQFT}
\end{align}
Define 
\begin{align}
R_{1,i}(t,s) & =M\{O_{1,i}P_{t}\}\text{, }R_{2,i}(t,s)=M\{O_{2,i}P_{t}\}.
\end{align}
then, from Eq.~(\ref{psO1}), we will obtain,
\begin{align}
\frac{\partial}{\partial s}R_{1,i} &=-i\omega_{i}R_{1,i}-i\lambda_{i}R_{1,i+1}-i\lambda_{i-1}R_{1,i-1}+\sum_{j}l_{i}^{*}l_{j}\int_{0}^{t}ds^{\prime}\alpha_{2}^{\ast}(s,s^{\prime})R_{2,j}^{\dagger}(t,s^{\prime})\nonumber \\
 &-\sum_{j}l_{i}^{*}l_{j}\int_{0}^{s}ds^{\prime}\alpha_{1}(s,s^{\prime})R_{1,j}(t,s^{\prime})-\sum_{j}l_{i}^{*}l_{j}\int_{s}^{t}ds^{\prime}\alpha_{2}(s^{\prime},s)R_{1,j}(t,s^{\prime}),\label{EqR1}
\end{align}
 with the initial condition $R_{1}(t,t)=a_{i}\rho$. In the derivation
above, the Novikov theorem 
\begin{align}
M\{w_{s}^{\ast}P_{t}\} & =\int_{0}^{t}ds^{\prime}\alpha_{2}^{\ast}(s,s^{\prime})M\{P_{t}O_{2}^{\dagger}(t,s^{\prime},z^{\ast},w^{\ast})\}\nonumber \\
 & =\sum_{i}l_{i}\int_{0}^{t}ds^{\prime}\alpha_{2}^{\ast}(s,s^{\prime})R_{2,i}^{\dagger}(t,s^{\prime}),
\end{align}
is used. Similarly, equation for $R_{2}$ can be derived as
\begin{align}
\frac{\partial}{\partial s}R_{2,i} & =i\omega_{i}R_{2,i}+i\lambda_{i-1}R_{2,i-1}+i\lambda_{i}R_{2,i+1}-\sum_{j}l_{i}l_{j}^{*}\int_{0}^{t}ds^{\prime}\alpha_{1}^{\ast}(s,s^{\prime})R_{1,j}^{\dagger}(t,s^{\prime})\nonumber \\
 &+\sum_{j}l_{i}l_{j}^{*}\int_{s}^{t}ds^{\prime}\alpha_{1}(s^{\prime},s)R_{2,j}(t,s^{\prime})+\sum_{j}l_{i}l_{j}^{*}\int_{0}^{s}ds^{\prime}\alpha_{2}(s,s^{\prime})R_{2,j}(t,s^{\prime}),\label{EqR2}
\end{align}
 with the initial condition $R_{2,i}(t,t)=a_{i}^{\dagger}\rho$. The
solutions for $R_{1}$, $R_{2}$ should take the following forms
\begin{align}
R_{1,i}&= \sum_{j}f_{i,j}(t,s)a_{j}\rho+\sum_{j}g_{i,j}(t,s)\rho a_{j},\label{R1}\\
R_{2,i}&= \sum_{j}u_{i,j}(t,s)a_{j}^{\dagger}\rho+\sum_{j}v_{i,j}(t,s)\rho a_{j}^{\dagger}.\label{R2}
\end{align}
 Substituting Eq.~(\ref{R1}-\ref{R2}) into Eq.~(\ref{EqR1}-\ref{EqR2}),
the equations for the coefficients can be derived as
\begin{align}
\frac{\partial}{\partial s}f_{i,j} & = -i\omega_{i}f_{i,j}-i\lambda_{i}f_{i+1,j}-i\lambda_{i-1}f_{i-1,j}+\sum_{k}l_{i}^{*}l_{k}\int_{0}^{t}ds^{\prime}\alpha_{2}^{\ast}(s,s^{\prime})v_{k,j}^{*}(t,s^{\prime})\nonumber \\
 &  -l_{i}^{*}l_{k}\int_{0}^{s}ds^{\prime}\alpha_{1}(s,s^{\prime})f_{k,j}(t,s^{\prime})-l_{i}^{*}l_{k}\int_{s}^{t}ds^{\prime}\alpha_{2}(s^{\prime},s)f_{k,j}(t,s^{\prime})
\end{align}
\begin{align}
\frac{\partial}{\partial s}g_{i,j} & = -i\omega_{i}g_{i,j}-i\lambda_{i}g_{i+1,j}-i\lambda_{i-1}g_{i-1,j}+\sum_{k}l_{i}^{*}l_{k}\int_{0}^{t}ds^{\prime}\alpha_{2}^{\ast}(s,s^{\prime})u_{k,j}^{*}(t,s^{\prime})\nonumber \\
 &  -l_{i}^{*}l_{k}\int_{0}^{s}ds^{\prime}\alpha_{1}(s,s^{\prime})g_{k,j}(t,s^{\prime})-l_{i}^{*}l_{k}\int_{s}^{t}ds^{\prime}\alpha_{2}(s^{\prime},s)g_{k,j}(t,s^{\prime})
\end{align}
\begin{align}
\frac{\partial}{\partial s}u_{i,j} & =i\omega_{i}u_{i,j}+i\lambda_{i-1}u_{i-1,j}+i\lambda_{i}u_{i+1,j}-\sum_{k}l_{i}l_{k}^{*}\int_{0}^{t}ds^{\prime}\alpha_{1}^{\ast}(s,s^{\prime})g_{k,j}^{*}(t,s^{\prime})\nonumber \\
 &  +l_{i}l_{k}^{*}\int_{s}^{t}ds^{\prime}\alpha_{1}(s^{\prime},s)u_{k,j}(t,s^{\prime})+l_{i}l_{k}^{*}\int_{0}^{s}ds^{\prime}\alpha_{2}(s,s^{\prime})u_{k,j}(t,s^{\prime})
\end{align}
\begin{align}
\frac{\partial}{\partial s}v_{i,j} & = i\omega_{i}v_{i,j}+i\lambda_{i-1}v_{i-1,j}+i\lambda_{i}v_{i+1,j}-\sum_{k}l_{i}l_{k}^{*}\int_{0}^{t}ds^{\prime}\alpha_{1}^{\ast}(s,s^{\prime})f_{k,j}^{*}(t,s^{\prime})\nonumber \\
 &  +l_{i}l_{k}^{*}\int_{s}^{t}ds^{\prime}\alpha_{1}(s^{\prime},s)v_{k,j}(t,s^{\prime})+l_{i}l_{k}^{*}\int_{0}^{s}ds^{\prime}\alpha_{2}(s,s^{\prime})v_{k,j}(t,s^{\prime})
\end{align}
 with the initial conditions $f_{i,j}(t,t)=u_{i,j}(t,t)=\delta_{ij}$,
$g_{i,j}(t,t)=v_{i,j}(t,t)=0$.

Then, substituting Eq.~(\ref{R1}-\ref{R2}) into Eq.~(\ref{MEQFT}),
the final master equation will be derived as it is shown in Eq.~(\ref{MEQFTfinal}).
Note that the coefficients $F$, $G$, $U$, $V$ are determined by
$f$, $g$, $u$, $v$ and the correlation functions $\alpha_{j}(t,s)$
($j=1,2$) as shown in Eq.~(\ref{FTF}).

\end{document}